\documentclass{mem}
\usepackage{natbib}\usepackage{txfonts}
\usepackage{graphicx}
\usepackage{epsf}
\usepackage{psfig}
\usepackage[a4paper]{hyperref}
\begin{document}

\title{
Driving in ZZ Ceti stars - Problem solved?
}

   \subtitle{}

\author{
A. Kim \, 
\inst{1} 
D.E. Winget \, 
\inst{1} 
M.H. Montgomery 
\inst{1} 
\, S.O. Kepler\inst{2}
          }
\institute{
The University of Texas at Austin,
Department of Astronomy,
Austin, TX 78712-1084, USA
\and
Instituto de F\'isica
Universidade Federal do Rio Grande do Sul
91501-900 Porto Alegre, RS, Brazil
\email{kepler@if.ufrgs.br}
}

\authorrunning{Kim et al. }

\titlerunning{Driving in ZZ Ceti stars}

\abstract{
There is a fairly tight correlation between the pulsation periods and effective temperatures
of ZZ Ceti stars (cooler stars have longer periods). This seems to fit the theoretical 
picture, where driving occurs in the partial ionization zone, which lies deeper and deeper
within the star as it cools. It is reasonable to assume that the pulsation periods should be 
related to the thermal timescale in the region where driving occurs. As that region sinks 
further down below the surface, that thermal timescale increases. Assuming this connection, 
the pulsation periods could provide an additional way to determine effective temperatures, 
independent of spectroscopy. We explore this idea and find that in practice, things are not 
so simple.
}

\maketitle{}

\section{Introduction}

ZZ Ceti stars, also called DAVs, are the coolest known pulsating white dwarfs (see figure 1).
They show only hydrogen in their spectra and are observed to pulsate in a narrow instability 
strip around 12000K. The number of known ZZ Ceti stars recently doubled 
(Mukadam 2004), and more have been found since (Mullally et al. 2005, Kepler et al. 2005, 
Castanheira et al. 2006, Castanheira 2006). With such a significant number of stars known, 
it was promising to study their ensemble characteristics in more depth (Mukadam 2004, 
Mukadam et al. 2005).

Even before the new ZZ Ceti's were discovered, a clear relation between their mean 
pulsating periods and spectroscopically determined effective temperatures was already 
apparent (Clemens 1993, Kanaan 1996). Cooler ZZ Ceti's show longer periods (see figure 3). 
Well aware of the dificulties associated with the spectroscopy of such stars, Mukadam et al. 
(2005) suggested that we could use this relation to obtain an independent measurement of 
temperatures for those stars.

We compared relevant thermal timescales of models with different effective temperatures. We 
looked in particular at driving through the $\kappa$-$\gamma$ mechanism (Cox \& Giuli 1968) 
and convective driving (Brickill 1991, Goldreich and Wu 1999). We expected that if we picked 
the right convective efficiency, the mean periods would agree with the thermal timescales (to
within a multiplicative constant). We would then use the effective temperatures of the models 
as a temperature scale.

Besides helping the ongoing effort to better define the ZZ Ceti instability strip, we 
then planned to apply the same method to the less numerous DBV stars (Helium 
atmospheres). Spectroscopic determinations of effective temperatures for those stars 
are difficult to do and uncertainties are large. We need more accurate
temperature determinations for DBV stars because they can be used to learn about 
plasmon neutrinos, and the results are most sensitive to temperatures 
(Winget et al. 2004, Kim et al. 2005, Kim et al. 2006)

\begin{figure}[!ht]
\resizebox{\hsize}{!}{\includegraphics[clip=true]{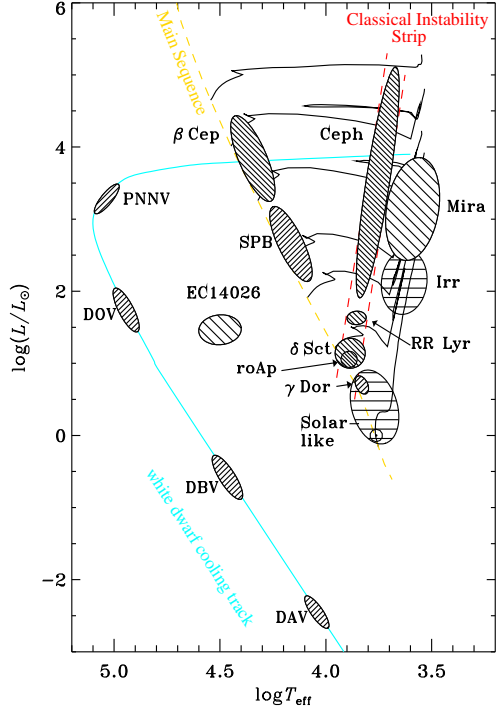}}
\caption{\footnotesize
H-R diagram showing the different classes of pulsating stars. In this article, we are
interested in the DAV's.
}

\end{figure}

\section{Pulsation driving and timescales}

Pulsations in ZZ Ceti stars are driven in the partial hydrogen ionization zone. We 
looked at two processes, each of which are associated with a thermal timescale. The
first process is the $\kappa$-$\gamma$ mechanism, and the relevant timescale in this case 
is the thermal timescale at the base of the partial ionization zone ($R_b$). Like 
Winget et al. (1982), we assume that only modes that have periods of the same order as 
the thermal timescale are driven. This thermal timescale is the time it takes 
to transport energy from $R_b$ to the surface of the star. It is equal to the heat 
content of the layers above $R_b$ divided by the total luminosity:

\begin{equation}
\tau_{th} \sim \frac{\Delta M_r c_V T}{L_{tot}}
\end{equation}

The second process is convective driving. Convective driving was proposed to attempt 
to resolve a discrepancy between the periods of the overstable modes and the above 
timescale. Noting that the convective turnover time was much shorter than the periods 
of the modes, Brickill (1991) assumed that convection responded instantanuously to the 
pulsations. Goldreich and Wu (1999) later revisited his theory of convective driving and 
arrived at a new thermal timescale, which is the time during which the convection zone
can bottle up the flow that enters from below. This new timescale $\tau_c$ differs 
from the traditional thermal timescale by a multiplicative constant roughly equal 
to 4.

So we have two timescales for which we know the dependence on the local temperature. To
compare with the observed periods, we need them in terms of effective temperature. 
Numerical calculations (Montgomery 2005) show that $\tau_c$, and therefore also $\tau_{th}$
are proportional to the effective temperature to a power roughly equal to -90 (for DAVs). This result is insensitive to the convective efficiency. Therefore in a 
$\log P-\log T_{\mathrm{eff}}$ plane, we 
would expect the data to lie on a line with a slope of -90, no matter which of the two
theories is the correct theory for driving in ZZ Ceti stars, and independently of the 
right convective efficiency.

\begin{figure}[!ht]
\resizebox{\hsize}{!}{\includegraphics[clip=true]{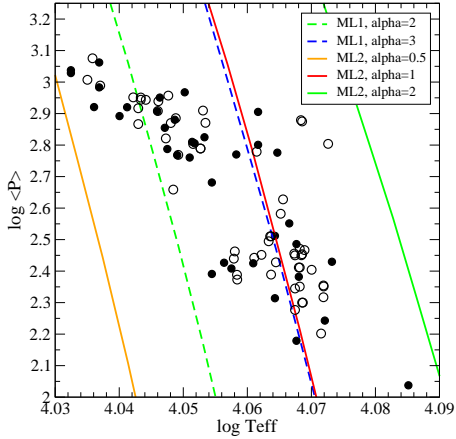}}
\caption{
\footnotesize
The ZZ Ceti stars discovered before Mukadam et al. 2004 (filled circles), and the ones
found by Mukadam et al. from SDSS data (open circles), along with $\tau_c$ for different 
convective efficiencies. In the legend, we designate B\"ohm-Vitense's treatment of 
convection (B\"ohm-Vitense 1958) by ML1, and the more efficient B\"ohm \& Cassinelli 
convection (B\"ohm \& Cassinelli 1971) by ML2. $\alpha$ is the ratio of the mixing length to
the pressure scale height. There is a systematic offset of $\sim 300K$ between the two 
samples, probably caused by systematic differences between Bergeron's models
atmospheres and Koester's, for which we artificially corrected by 
substracting 294 K from the ``old'' sample. The choice is somewhat arbitrary since we 
are only concerned with relative temperatures at this point. The three open circles at 
high effective temperature and high $<P>$ that seem to deviate from the correlation also 
happen to have high gravities. This is consistent with theory, which predicts that 
$\tau_c$ is significantly higher for high gravities (about an order of magnitude for 
$\rm \Delta log(g) = 0.6$)
}
\end{figure}

\begin{figure}[!h]
\resizebox{\hsize}{!}{\includegraphics[clip=true]{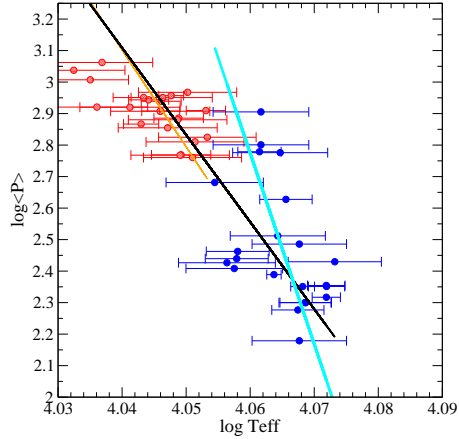}}
\caption{
\footnotesize
A subset of the ZZ Ceti stars (those with $\log g$ between 8.0 and 8.2) and different fits.
The black line is a fit to the whole subset (slope = -28), the thin grey (orange) line is 
a fit to the cool end stars (slope = -31), and the thick grey (light blue) line is a fit 
to the hot end stars (slope = -61). The boundary between the cool (cDAVs) and hot stars 
(hDAVs) is 11320 K.
}
\end{figure}

\section{Results}

Figure 2 shows the data on a $\log P - \log T_{\mathrm{eff}}$ plane along with curves that 
show $\tau_c(T_{\mathrm{eff}})$ for different convective efficiencies. We see that indeed, 
the thermal timescale curves are parallel to each other, and we also see that they have a 
different slope than the data.

Figure 3 shows a subset of the data with error bars. We have retained stars in a 
narrower $\log g$ range (8.0 to 8.2) in a rudimentary attempt to remove the $\log g$ 
dependence of the timescales. We fit one line to the cool DAVs, one to the hot DAVs,
and a third to the full effective temperature range. The boundary is at 11320K. 
The slopes of the three lines are, respectively, -31, -61, and -28. Uncertainties are 
large and those results are not entirely incompatible with the expected slope of -90.
The difference in slope between the red-edge pulsators and the blue-edge pulsators is
not statistically significant (again because of the uncertainties)
but theoretically the blue-edge pulsators don't show significant convection zone.

We do note, however, that departures from the models become important for cooler 
stars. It looks like a saturation mechanism is at work, which prevents modes much 
higher than 1000s from being driven, though that is not the only explanation.

\section{Some ideas}

In our analysis, we have made a few assumptions. First we assumed that the convective 
efficiency remains the same as the stars cools. Maybe convection becomes less 
effective with decreasing effective temperature, which would explain the observed 
shorter timescales.

While this may seem to go against common wisdom, there are a few physical processes 
which could work against convection for cooler stars. Magnetic fields may become more
important as the star cools and the depth of the convection zone increases. One could
easily imagine the magnetic field lines constraining the motion of the plasma and 
inhibiting the convection, leading to a lower convective efficiency.

Most red edge ZZ Ceti stars show large amplitude pulsations. The linear approximations
break down, and one could easily imagine that such high amplitude pulsations could 
have an effect on the convective efficiency. At the red edge, the pulsation amplitudes
can be as high as 20\% at the surface (Kleinman 1995). At those amplitudes, the energy 
carried by the pulsations is non-negligible, which would lead to a less efficient
convection. This of course fails to explain why the red edge ZZ Ceti's that are not 
high amplitude pulsators deviate from the theoretical thermal timescales in the same 
way the others do.

Maybe it is our assumption that the weighted mean period (i.e. the period(s) of the 
dominant mode(s)) is $\sim \tau_{c}$ that is flawed. Again, this may not work so well 
where non-linearities are important. At lower effective temperature, the convection zone 
grows deeper, and when deep enough, maybe its motion affects the timescales.

The most obvious explanation, and maybe the best one we have is that modes with 
periods higher than about 1200 seconds are suppressed through a different mechanism.
Lack of surface reflection comes to mind (Hansen et al. 1985). Modes with long wavelengths 
(long periods) are not reflected very efficiently in
the outer layers of the stars where the density falls off exponentially, and lose 
their energy. This places an upper limit on the periods of the driven modes. Hansen et al. 
(1985) did a simple non-adiabatic calculation and found that modes with periods
greater than 3000s could not sustain themselves in ZZ Cetis, although it is possible that
The actual limit may be more like 1200s.

\section{Conclusion}
We set out to develop an independent method to determine temperatures for ZZ Ceti 
stars, with the intention to apply the same method later to DBV stars. We based our 
analysis on the assumtpion that the period(s) of the dominant mode(s) should be at 
least proportional to timescales corresponding to several theories of driving in 
ZZ Ceti stars. In actually doing the calculations, we found that this was not the case.
While the theory predicts that in the $\rm log<P>-\;logT_{eff}$ plane, the data should lie
on a line with slope $\sim -90$. We find the actual slope to be less steep.
 
We give a few ideas to explain the discrepancy, but this remains somewhat of a mystery, 
which would be very interesting to elucidate in order to not only obtain better effective 
temperatures, but also to improve our understanding of driving in ZZ Ceti stars.

\bibliographystyle{aa}

\end{document}